\documentclass[aps,pre,groupedaddress,amsmath,amssymb,twocolumn]{revtex4-1}

\usepackage[T1]{fontenc} 
\usepackage[utf8]{inputenc}
\usepackage{graphicx}
\usepackage{subcaption}
\usepackage{nameref,hyperref}

\date{\today}

\usepackage{graphicx}

\begin{document}

\title{A Minimal Model of Burst-Noise Induced Bistability} 

\author{Johannes Falk}
 \email[]{falk@fkp.tu-darmstadt.de} 
\author{Marc Mendler}
 \email[]{marcm@fkp.tu-darmstadt.de} 
\author{Barbara Drossel}
 \email[]{drossel@fkp.tu-darmstadt.de}

 \affiliation{Institut für Festkörperphysik, Technische Universität Darmstadt, Hochschulstr. 6, 64289 Darmstadt, Germany}

\begin{abstract}
We investigate the influence of intrinsic noise on stable states of a one-dimensional dynamical
system that shows in its deterministic version a saddle-node bifurcation between monostable and bistable behaviour. The system is a modified version of the Schl\"ogl model, which is a chemical reaction system with only one type of molecule.
The strength of the intrinsic noise is varied without changing the deterministic description by introducing bursts in the autocatalytic production step. We study the transitions between monostable and bistable behavior in this system by evaluating the number of maxima of the stationary probability distribution. We find that changing the size of bursts can destroy and even induce saddle-node bifurcations. This means that a bursty production of molecules can
qualitatively change the dynamics of a chemical reaction system even when the deterministic description remains unchanged. 
\end{abstract}

\maketitle

\section{Introduction}

Many dynamical systems in nature and technology have an underlying discrete and
stochastic behaviour. Examples are population growth
\cite{kendall_stochastic_1949}, chemical reaction systems
\cite{gillespie_exact_1977}, metabolic and gene transcription dynamics in biological cells, or free-way traffic
\cite{nagel_cellular_1992}. For a long time, the description of these systems was mainly done using deterministic equations for the average concentrations of molecules or individuals, which is a good approximation when numbers are large. 
The analysis of such deterministic systems like for example the investigation and characterization of their stable states is well understood, and many analytical as well as numerical tools are available \cite{guckenheimer_nonlinear_2002}. In order to take into account the influence of the environment, additive or multiplicative noise was added to the deterministic equations. External influences were assumed to be the only
important source of noise \cite{horsthemke_noise_1984,du_stochastic_2016}.

However, theoretical calculations and simulations show that for many systems there  exists a significant difference between the simplified deterministic behavior and that obtained using the full information about the stochastic nature of the microscopic dynamics of the system, especially if the copy numbers of molecules or individuals are small\cite{durrett_importance_1994}. The origin of stochastic fluctuations lies in the discrete nature of the reactions,  and hence this type of noise is purely intrinsic. In particular, stochastic noise plays a crucial role in the dynamics of biological systems on the single-cell level \cite{horsthemke_noise_1984}. Until recently, it was however not possible to observe the predicted stochastic effects in biological cells due to a lack of appropriate measurement techniques in Micro and Systems Biology.  The development of a large pool of new techniques in recent years made it possible to study growth and metabolism at the single cell level. Examples are fluorescence techniques combined for instance with flow cytometry or microfluidic cell culture analysis \cite{nebe-von-caron_analysis_2000, B515632G}.

The effects of intrinsic noise on the dynamics of a system are manifold \cite{mckane_2014}. While noise was always considered to increase variations and fluctuations, it now 
becomes evident that it can also stabilize dynamical systems and 
de facto improve the signal e.g. by Stochastic Focusing \cite{paulsson_random_2000, paulsson_stochastic_2000, wiesenfeld_stochastic_1995}. 

One important variety of intrinsic noise is the so-called burst noise, where a single transition changes the number of individuals by at least two. Examples for burst noise are the bursty productions in gene transcription and translation \cite{raj_stochastic_2008,friedman_linking_2006}, or the simultaneous production of several offspring in litters of mammals \cite{humphries_determinants_2000}. Several studies have shown that bursts increase the impact of noise and can dramatically change the behavior of a stochastic system \cite{bokes_gene_2016, tsimring_noise_2014}. 

When the transition rates of such a stochastic system do not depend on the past but only on the present state, its dynamics is correctly described by a Master Equation. Unfortunately, in most cases the solution of the Master Equation is computationally demanding and analytically intractable. For this reason, several analytical tools have been developed to capture the effect of intrinsic noise in a simplified description and to study its effect on the dynamics and stability of a system. One often-used approach is the van Kampen system-size expansion, which exploits the fact that for larger particle numbers fluctuations are small compared to the mean value of the distribution. While the leading order of this expansion simply results in a Gaussian stationary distribution that is centered at the equilibria of the deterministic equation, expansion to the second order \cite{grima_effective_2010} yields correction terms to the deterministic equations that shift the equilibrium points. Similar results have been obtained by e.g. \cite{gomez-uribe_mass_2007} using moment closure techniques on the Master Equation. 
Using similar expansions, M.~Scott evaluates the time scale of the decay of the autocorrelation function to characterize the stability of stable states in stochastic systems \cite{scott_deterministic_2007}. One important finding is that increasing noise -- and especially burst noise -- can destabilize a stable point. 
While these and similar studies are tailored at describing stochastic systems in the neighborhood of (isolated) fixed points of the deterministic equations, different tools are required for evaluating the effect of intrinsic noise on bifurcations and on the existence of fixed points. There are several examples of systems with bifurcations and bistable behavior that have no counterpart in the macroscopic, deterministic description. Utilizing a continuous master equation, Friedman et al. \cite{friedman_linking_2006}  showed for example that noise introduces bistability in a gene expression network with self regulating transcription factors, even though the deterministic system is monostable. Several other authors showed similar effects in different systems with multiple species \cite{samoilov_stochastic_2005,mcsweeney_stochastically-induced_2014, bishop_stochastic_2010,karmakar_positive_2007,duncan_noise-induced_2015,assaf_determining_2011}. 

In this paper, we will deal with bistability that is induced by intrinsic noise in an even simpler system. This system has only one species and uses mass-action kinetics. This system is a simplified version of the Schl\"ogl model
\cite{schlogl_chemical_1972}, which was originally introduced to study
nonequilibrium phase transitions \cite{grassberger_phase_1982,erban_analysis_2008,vellela_stochastic_2009,gaspard_fluctuation_2004}. Due to its relevance for phase transitions, the characterization of bifurcations in the system has attracted researchers from various fields.  Because of its simplicity, the Schl\"ogl model can be used as a
generic template for a full class of one-dimensional bistable systems
\cite{wilhelm_smallest_2009}. The simplicity is achieved at the expense of introducing a tri-molecular reaction, for which reason the model was long considered to be only a theoretical concept. Nevertheless recently a first group succeeded at mapping a biological system onto the Schl\"ogl model \cite{endres_bistability:_2015}. So far, this model was mostly used in its deterministic noise-free version, where it shows saddle-node bifurcations as well as SNIPER bifurcations (Saddle-Node Infinite-Periodic bifurcations)\cite{erban_analysis_2008}.  

In the following, we will study a stochastic version of this model that includes burst noise. 
 Our analysis shows for this one-dimensional system that
on the one hand burst noise can destroy bistability, and on the other hand that it can also induce bistability. This effect is barely visible in the conventional stochastic version of the Schl\"ogl model, but becomes rather pronounced in the presence of burst noise. Our analysis is based on the Fokker-Planck-Equation (FPE) that is a good approximation based on a  continuous probability distribution the dynamics of which is essentially captured by a deterministic drift term and a noise-driven diffusion term \cite{gardiner_stochastic_2009}. We will evaluate the extrema of the stationary FPE (\ref{eq:fokkersolv}) to obtain the
stable states of the system. This way we show for the first time how 
intrinsic noise can induce a bistable behavior in a one-dimensional model.
Given the biological importance of the bistable behaviour of dynamical systems and due to its simplicity our approach provides several starting points for detailed investigations and can easily trigger further research.

The remainder of this paper is organized as follows. In the next two sections we introduce the model and the methods and analyze the conventional Schl\"ogl model. In the following section we
add bursting noise to the system and demonstrate how this induces and destroys bistability.  In the last section
 we draw some conclusions.

\section{Model \label{sec:model}}

\subsection{Schl\"ogl model}

We analyze a simplified version of the Schl\"ogl model,
which is considered to be the simplest possible one-dimensional
bistable system \cite{wilhelm_smallest_2009}. Its 
chemical reactions are
\begin{align}
 \label{eq:schloegl}
    \emptyset &\xrightarrow[]{k_1} X \nonumber \\
  X &\xrightarrow[]{k_2} \emptyset \nonumber  \\
  2X &\xrightarrow[]{k_3} 3X \\
  3X &\xrightarrow[]{k_4} 2X \nonumber
 \end{align} 
Later, we will study a modification of the third reaction,
\begin{equation}
 2X \xrightarrow[]{k_3/r} (2+r)X ,  \tag{1'} \label{burst}
\end{equation}
which represents a bursting production of $r$ molecules $X$ at the same time. Note that in this notation a burst size of $r=1$ corresponds to the unmodified system.
\subsection{Three different levels of mathematical modelling}
\subsubsection{Deterministic ODE}
Denoting the time-dependent number of molecules of substance $X$ 
with $x$, the deterministic  ODE of the Schl\"ogl model is 
\begin{align}
  \label{eq:detschloegl}
  \frac{dx}{dt} = k_1 - k_2 x + k_3 x^2 - k_4 x^3\, .
\end{align}
Using the propensity vector $\nu = [k_1,k_2 x
  ,k_3x^2,k_4 x^3]$ and the stoichiometric matrix $\mathbf{S} =
[1,-1,1,-1]$, the right-hand side can be written as a product,
\begin{align}
  \label{eq:snu}
  \frac{dx}{dt} = \mathbf{S} \nu \, .
\end{align}
The two factors on the right-hand side have an intuitive meaning. The
propensity vector indicates how often each reaction occurs, the
stoichiometric matrix determines how each reaction changes the number of molecules.

The deterministic ODE is a good description of the system if molecule numbers are sufficiently large that stochastic fluctuations can be neglected.  
\subsubsection{Master-Equation}
When stochastic effects are relevant, the reactions \eqref{eq:schloegl} are  translated into  the chemical master
equation (CME) \cite{vanKampen2007ix}, which gives the time evolution  of the probability $P(x,t)$ of having $x$ molecules at time $t$,
\begin{align}
\label{eq:mastereq}
\frac {\partial P(x,t)}{\partial t} = \sum_{j=1}^{R}\left(
\mathbb{E}^{-S_{j}} -1 \right) \nu_j P(x,t)\, .
\end{align}
Here, $\mathbb{E}^x$ is the step operator, which acts on a function $f(x)$ according to
\begin{align}
\mathbb{E}^a f(x) = f(x+a)\, ,
\end{align}
and $R$ is the number of reactions.
For the set of reactions \eqref{eq:schloegl}, equation \eqref{eq:mastereq} becomes:
\begin{align}
\frac{\partial P(x,t)}{\partial t} = 
\left(\mathbb{E}^{1} -1 \right) \left( T(x-1|x) P(x,t) \right) + \\
\left(\mathbb{E}^{-1} -1 \right) \left( T(x+1|x) P(x,t) \right)\, , \nonumber
\end{align}
where
\begin{align}
T(x+1|x) = k_1 + k_3 x^2  \, , \\
T(x-1|x) = k_2 x + k_4 x^3\, . \nonumber
\end{align}
Under the assumption
of a well stirred, thermally equilibrated system the CME can be shown to be exact
\cite{gillespie1992rigorous}.  The CME can rarely be analytically solved, and therefore computer simulations based on the Gillespie algorithm (SSA)
\cite{gillespie_exact_1977} are used to approximate sample
trajectories $x(t)$ of the system.  Since one needs a huge number of
trajectories to obtain the full distribution $P(x,t)$, the CME approach requires high computational effort.

\subsubsection{Fokker-Planck equation}
In order to proceed with analytical calculations, the CME is approximated by a Fokker-Planck equation (FPE), which is obtained by performing a  Taylor
expansion up to the second order in the changes $\Delta x$ that occur during a small time $dt$,  as was first done by
Kramers and Moyal \cite{gardiner_stochastic_2009}. The first order term of this expansion is the "drift term", which describes the deterministic change in the absence of noise, just as the ODE \eqref{eq:detschloegl}. The second order term is the "diffusion term", which describes the increase of the width of the distribution. The general form of the FPE is
\begin{align}
  \label{eq:fokker}
  \frac{\partial P(x,t)}{\partial t} &=  \\ 
  & -  \frac{\partial}{\partial x} A(x) P(x,t)\, + \notag \\
  &\frac{1}{2}  \frac{\partial^2}{\partial x^2} B(x) P(x,t)\, , \notag
\end{align}
where $A$ is the first moment of $\Delta x$, and $B$ the second moment, divided by $dt$. 
For our model, we have 
\begin{align}
  \label{eq:drift}
  A(x) & = k_1 - k_2 x + k_3 x^2 - k_4 x^3 = \mathbf{S} \nu \nonumber\\ B (x)& =
  \left( k_1 + k_2 x + k_3 x^2 + k_4 x^3\right) =
  \mathbf{S}~diag(\nu)~\mathbf{S}^{-1}\, .
\end{align}
In the limit $t \rightarrow \infty$ Eq.~(\ref{eq:fokker}) becomes stationary and independent of time,
hence we can write:
\begin{align}
  \frac{\partial}{\partial x} A(x) P_s(x) = \frac{1}{2}
  \frac{\partial^2}{\partial x^2} B(x) P_s(x)\, ,
\end{align}
which is -- assuming realistic boundary conditions -- solved by
\cite{gardiner_stochastic_2009}
\begin{align}
  \label{eq:fokkersolv}
  P_s(x) = \frac{\mathcal{N}}{B(x)} \exp \left[ \int_0^x
    \frac{2 A(x')}{B(x')} dx'\right] ,
\end{align}
where $\mathcal{N}$ is a constant that normalizes the distribution. It
can be shown that this solution is exact for systems with 
Gaussian white noise \cite{hanggi_stochastic_1982,risken_fokker-planck_1984}.
  
\section{Stationary Solutions and bifurcation diagrams of the Schl\"ogl model}
\label{sec:detstat}
Before studying the influence of bursting noise, let us briefly summarize the properties of the unmodified model.    
The stationary solution of the deterministic ODE is obtained by setting $A(x)=0$, giving three roots $x_1$, $x_2$, and $x_3$. The condition that all roots are real  is given by
\begin{align}
 \label{eq:disc}
\Delta \equiv& -27 \tilde k_1^2+18 \tilde k_1 \tilde k_2 \tilde k_3 - 
\\ \notag
&\quad 4 \tilde k_1 \tilde k_3^3-4 \tilde k_2^3+ \tilde k_2^2 \tilde k_3^2 > 0
\end{align}
with $\tilde k_i=k_k/k_4$. Hence for $\Delta >0$ the system is bistable, with one unstable fixed point between the two stable fixed points, for $\Delta < 0$ it is monostable \cite{irving_integers_2003}. The transition from the monostable to the bistable regime occurs via a saddle-node bifurcation. 

\begin{figure}
   \includegraphics[width=0.4 \textwidth]{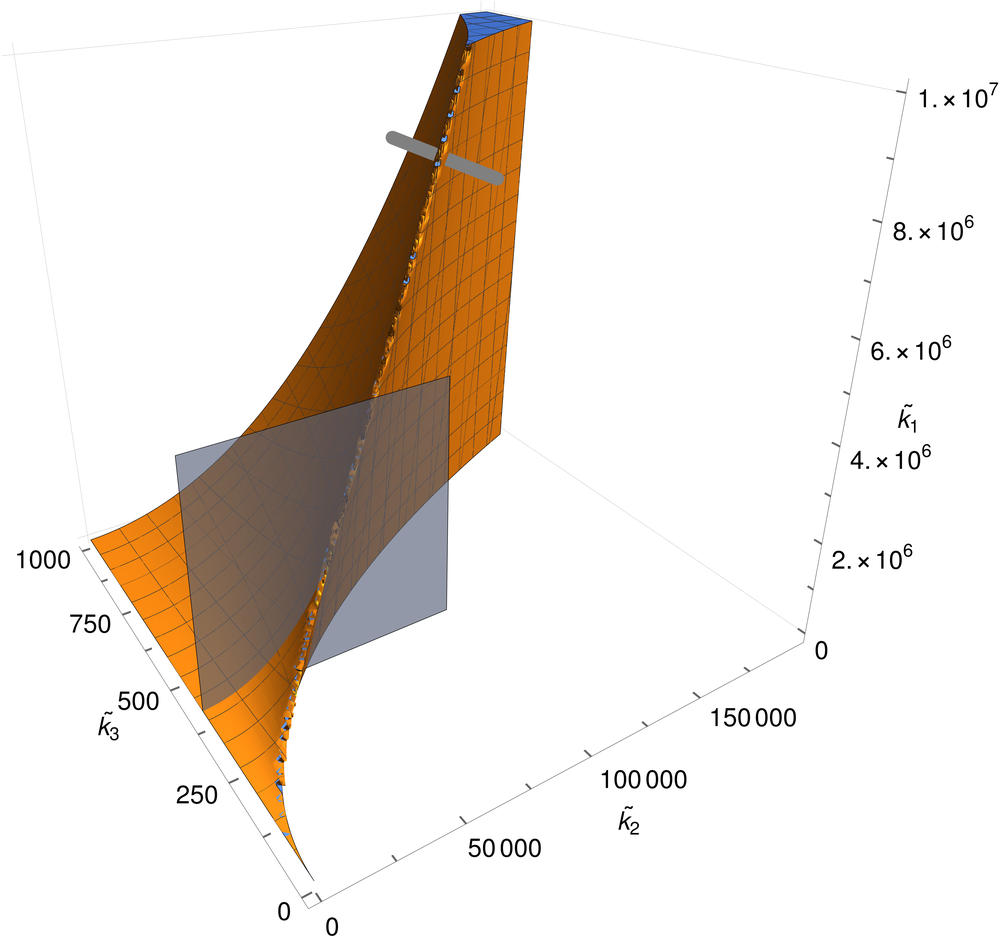}\\
    \includegraphics[width=0.4 \textwidth]{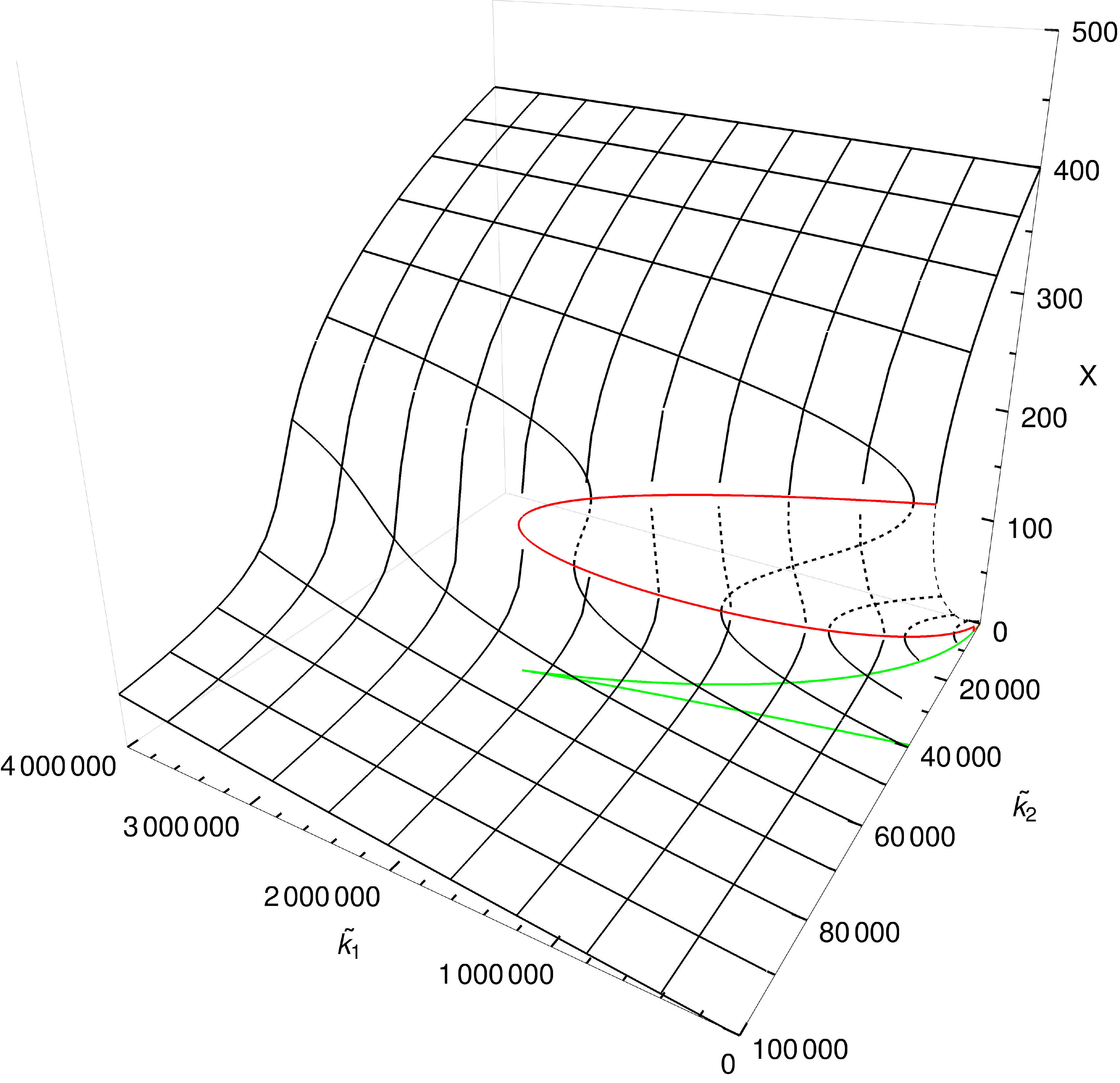}
   \caption{
Top: The bistable (filled) and monostable parameter regions
      of the deterministic model. Bottom: Steady-state molecule numbers for a two-dimensional cross section, showing the two saddle-node bifurcations that merge, creating a cusp bifurcation. The red line indicates the transitions from stable to unstable. The green line is the projection of the red line onto the $k1-k2$-plane. The gray plane in the top graph visualizes the region that is plotted in the bottom figure. 
  The gray line in the top graph is the trajectory of parameters later used to produce Fig. \ref{fig:fokker_compare}
    \label{fig:number_of_stable}}
       \end{figure}

\begin{figure}
   \includegraphics[width=0.4 \textwidth]{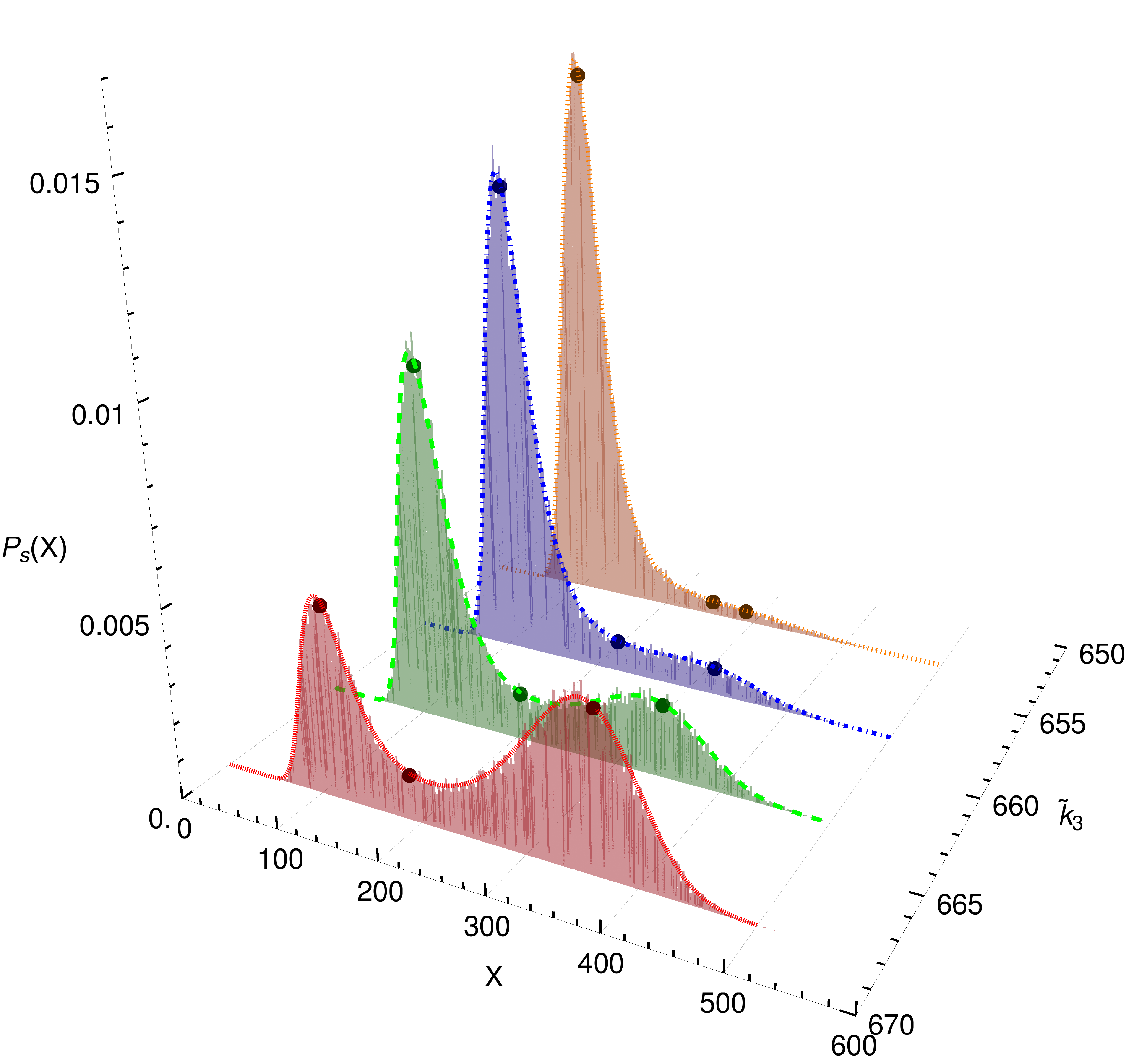}
     \caption{Stationary distributions obtained by stochastic simulation (histograms)     
 and by solving the Fokker-Planck-Equation (lines) for
     different values of $\tilde{k}_3$ (other parameters: $\tilde{k}_1 = 8\cdot 10^6, \tilde{k}_2 = 1.33 \cdot 10^5$, see the gray line in Fig.~\ref{fig:number_of_stable}). The dark dots indicate the fixed points of the deterministic model. \label{fig:fokker_compare}}%
 \end{figure}

Fig. \ref{fig:number_of_stable} shows the bistable and monostable parameter regions in three-dimensional parameter space, and the stationary solution $x$ in a two-dimensional cross section.   
When noise is taken into account, the stationary solutions are not points but distributions that have one or two local maxima that are not exactly at the location of the deterministic fixed points, see Fig.~\ref{fig:fokker_compare}. 
The stochastic simulations for this figure were performed with the Gillespie
algorithm using the free software package
\textit{Dizzy} \cite{ramsey_dizzy_200}. They agree very well with the stationary solution of the FPE, which was obtained analytically from  Eq.~(\ref{eq:fokkersolv}) by writing eq.~(\ref{eq:drift}) in the form
\begin{align}
  A(x) = -k_4 (x-x_1)(x-x_2)(x-x_3)
\end{align}
and
\begin{align}
  B(x) = k_4 (x + x_1)(x + x_2)(x + x_3)
\end{align}
and performing a partial fraction decomposition, giving  
\begin{align}
  \label{eq:fokkerfin1}
\begin{split}
  P_s(X)= \mathcal{N} e^{2x} (x_1 + x)^{\frac{2 x_1 (x_1 + x_2) (x_1 +
      x_3)}{(x_1 - x_2)(x_1 - x_3)}-1} \\ (x_2 + x)^{\frac{2 x_2 (x_2
      + x_1) (x_2 + x_3)}{(x_2 - x_1)(x_2 - x_3)}-1}\\ (x_3 +
  x)^{\frac{2 x_3 (x_3 + x_1) (x_3 + x_2)}{(x_3 - x_1)(x_3 - x_2)}-1}\, .
\end{split}
\end{align}
The maxima and minima of $P(x)$ can be obtained directly from 
Eq.~(\ref{eq:fokkersolv})\cite{gillespie1991markov},
\begin{align}
  \alpha(x) := A(x) - \frac{1}{2} B'(x) = 0\, .
  \label{eq:mes_ext}
\end{align}
All bifurcations shown in Fig. \ref{fig:number_of_stable} and Fig. \ref{fig:fokker_compare} have been generated by changing one of the
parameters $\tilde{k}_i$ and hence changing the deterministic equation. In the following, we will show that it is possible to obtain bifurcations without changing the deterministic equations of the
model. This is done by changing the temporal pattern of reaction rates (by introducing reaction bursts) that do not change the drift term but only the diffusion term of the Fokker-Planck equation.
 
\section{Influence of Bursting Noise \label{sec:bnoise}}
  Now we investigate the modified Schl\"ogl model, using reaction (\ref{burst}).
The propensity vector and the stoichiometric matrix then become
\begin{align}
  \nu &= [k_1,k_2 x ,\frac{k_3 x^2}{r},k_4 x^3]^T \\ \mathbf{S} &=
      [1,-1,r,-1]
\end{align}
The deterministic ODE,  $\frac{dx}{dt}=A(X)= \mathbf{S} \nu$ is not changed by the bursting noise, because the reaction (\ref{burst}) occurs $r$ times less often when it produces $r$ times as many molecules. The strength of the intrinsic noise is obviously increased with increasing $r$. The functions $A$ and $B$ occuring in the Fokker-Planck equation become now
\begin{align}
  \label{eq:ab_burst}
  A &= k_1 - k_2 x + k_3 x^2 - k_4 x^3 = \mathbf{S} \nu \\ B &=
  \left( k_1 + k_2 x + r k_3 x^2 + k_4 x^3 \right) =
  \mathbf{S}~diag[\nu]~\mathbf{S}^{-1}
\end{align}
Since $A$ and $B$ now have different roots, Eq.~(\ref{eq:fokkerfin1})
is no longer the valid solution.

\begin{figure}
   \includegraphics[width=0.4 \textwidth]{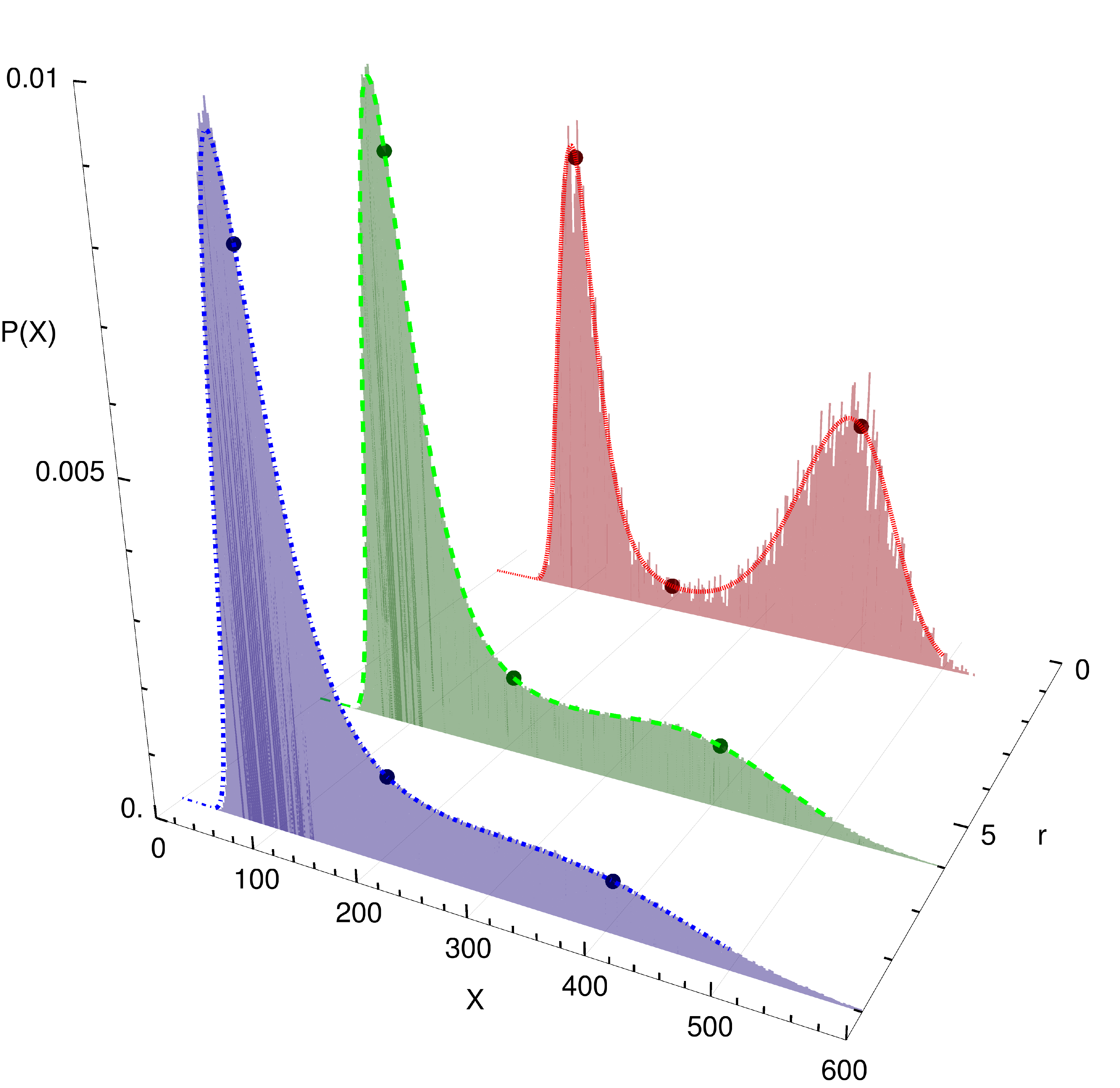}
     \caption{
Stationary distributions obtained by stochastic simulation (histograms)     
 and by solving the Fokker-Planck-Equation (lines) for
     different values of the burst size $r$ (other parameters: $\tilde{k}_1 = 9 \cdot 10^6, \tilde{k}_2 = 1.5 \cdot 10^5, \tilde{k}_3 = 720$). The dark dots indicate the fixed points of the deterministic model, which do not depend on $r$. \label{fig:mes_compare}}%
 \end{figure}

\begin{figure}
   \begin{subfigure}[b]{0.3\textwidth}
     \centering
     \includegraphics[width=\textwidth]{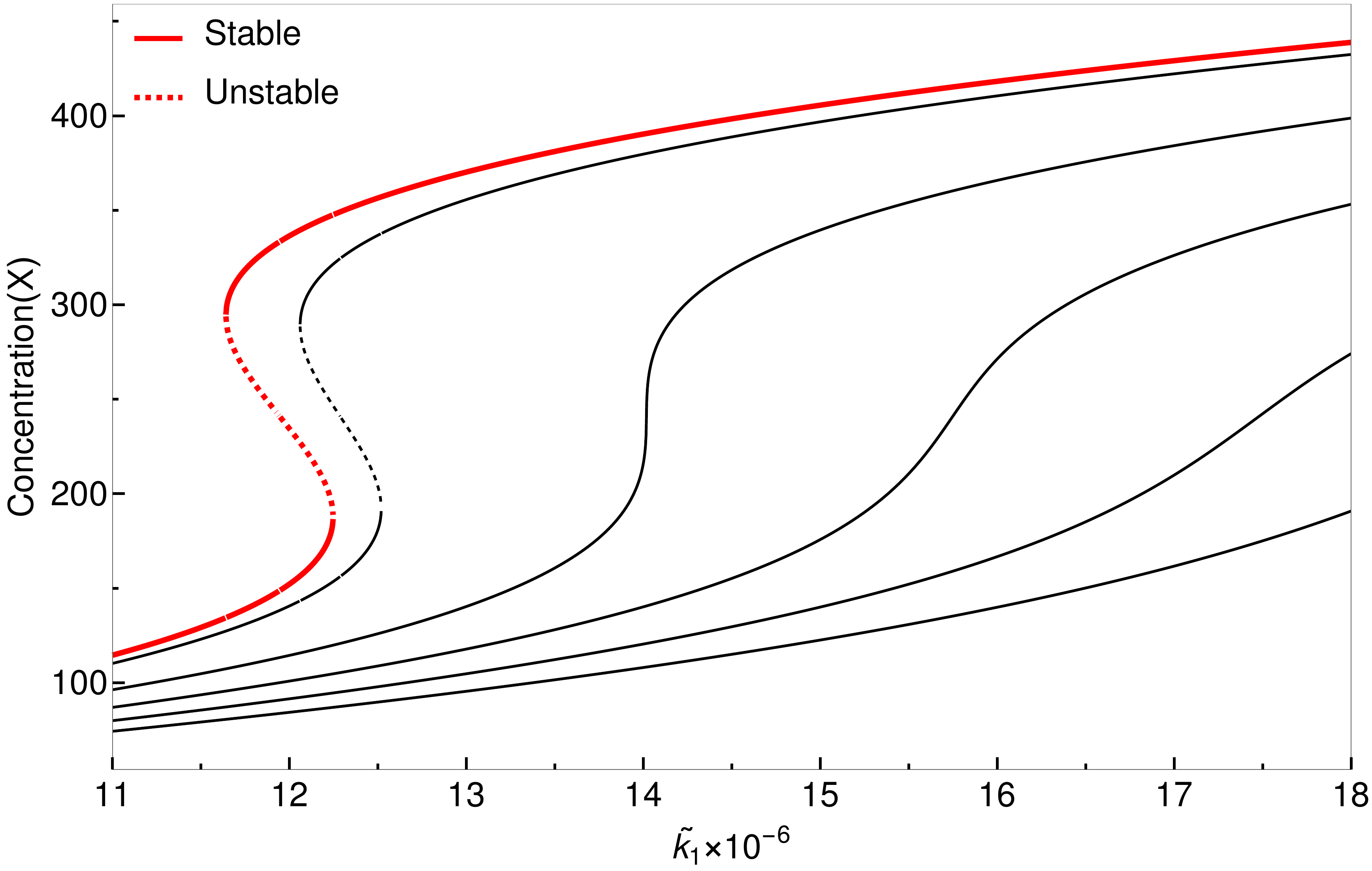}
     \caption{}
   \end{subfigure}
   \begin{subfigure}[b]{0.3\textwidth}
     \centering
     \includegraphics[width=\textwidth]{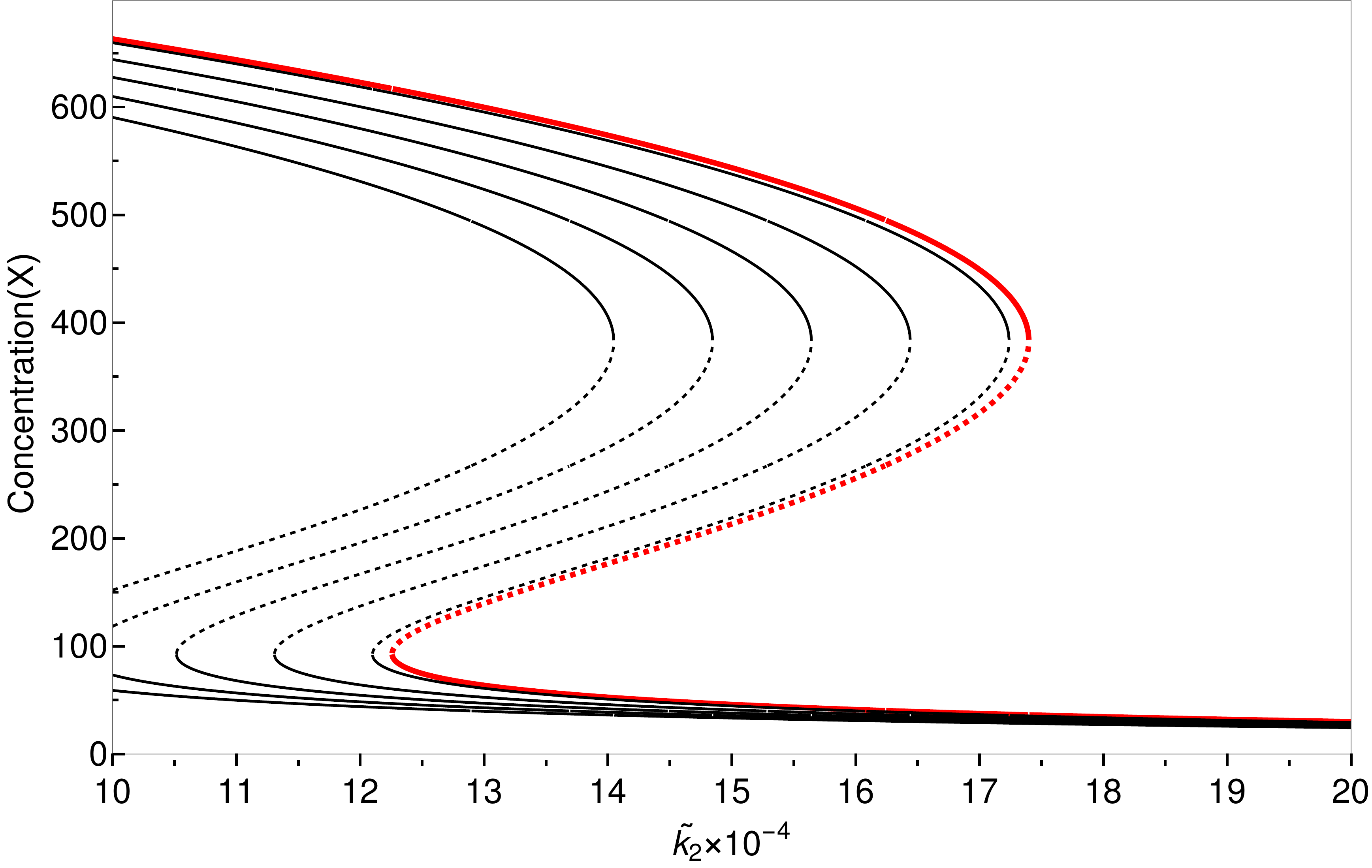}
     \caption{}
   \end{subfigure}
      \begin{subfigure}[b]{0.3\textwidth}
     \centering
     \includegraphics[width=\textwidth]{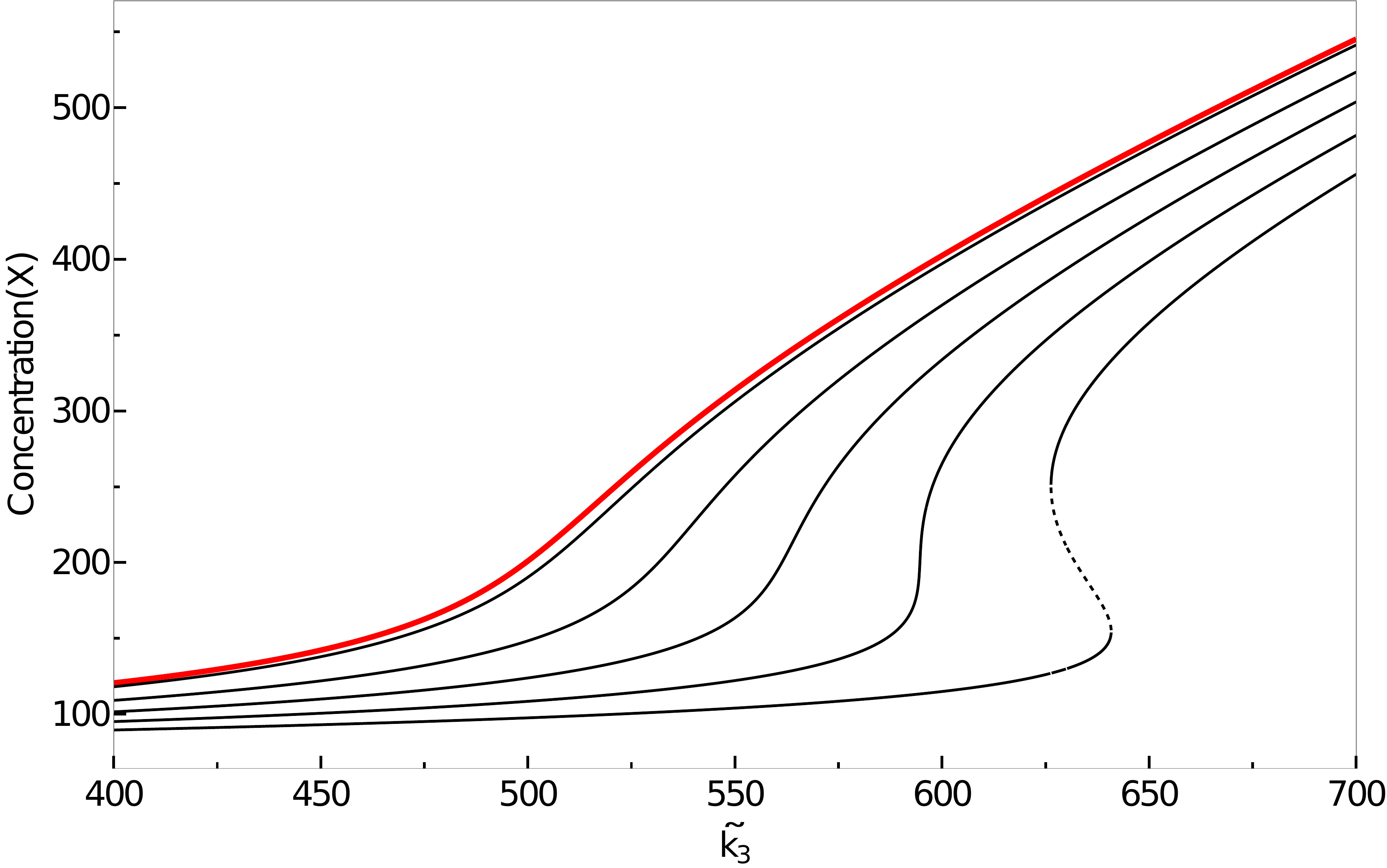}
     \caption{}
   \end{subfigure}
      \caption{
Influence of noise on the maxima of the stationary solution, showing that noise shifts the maxima to the extent that bifurcations can be destroyed or induced. The red line shows the deterministic fixed points, the first black line shows the solution for a burst size 1 (the standard Schl\"ogl model). With each further black line, the burst size $r$ increases by 10. Starting from the first black curve next to the red one, the parameter value $r$ of the black curves are therefore: 1,11,21,31,41. The other parameters are (a) $\tilde{k}_2 = 1.65\cdot 10^5, \tilde{k}_3 = 720$, (b) $\tilde{k}_1 = 5.33 \cdot 10^{6}, \tilde{k}_3 = 800$, (c) $\tilde{k}_1 = 8 \cdot 10^{6}, \tilde{k}_2 = 1\cdot 10^5$.  
 \label{fig:det_compare}}
 \end{figure}

\begin{figure}
 \includegraphics[width=0.45 \textwidth]{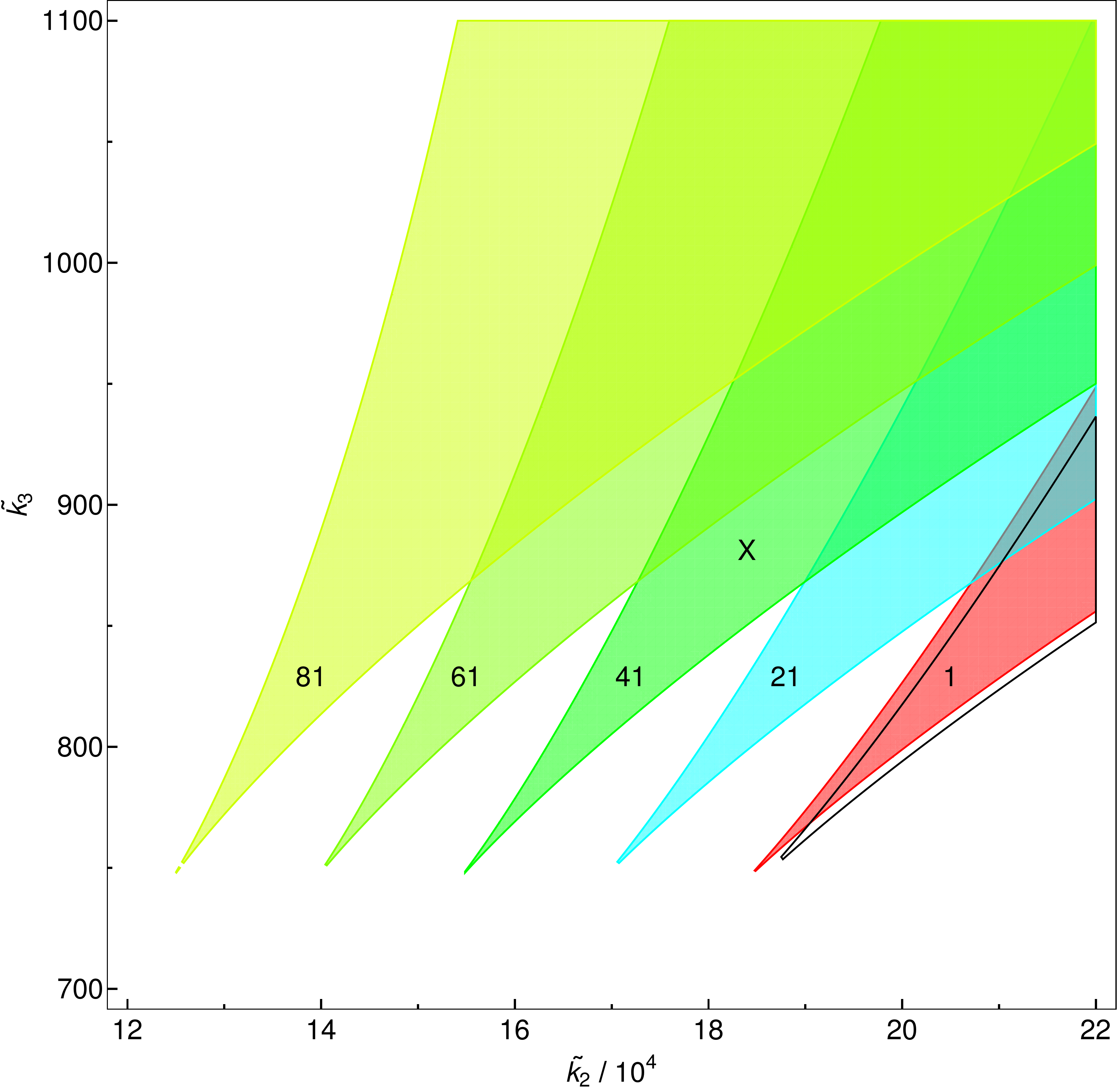}
 \caption{
Cross section of the phase diagram in the $\tilde{k}_1 = 1.54$ plane. The burst size for the red surface is $r=1$ for each step to the left $r$ is increased by 20 (as indicated by numbers in each colored tongue). The black line shows the deterministic solution. \label{fig:crosssect_burst}}%
 \end{figure}

\begin{figure}
\includegraphics[width=0.45 \textwidth]{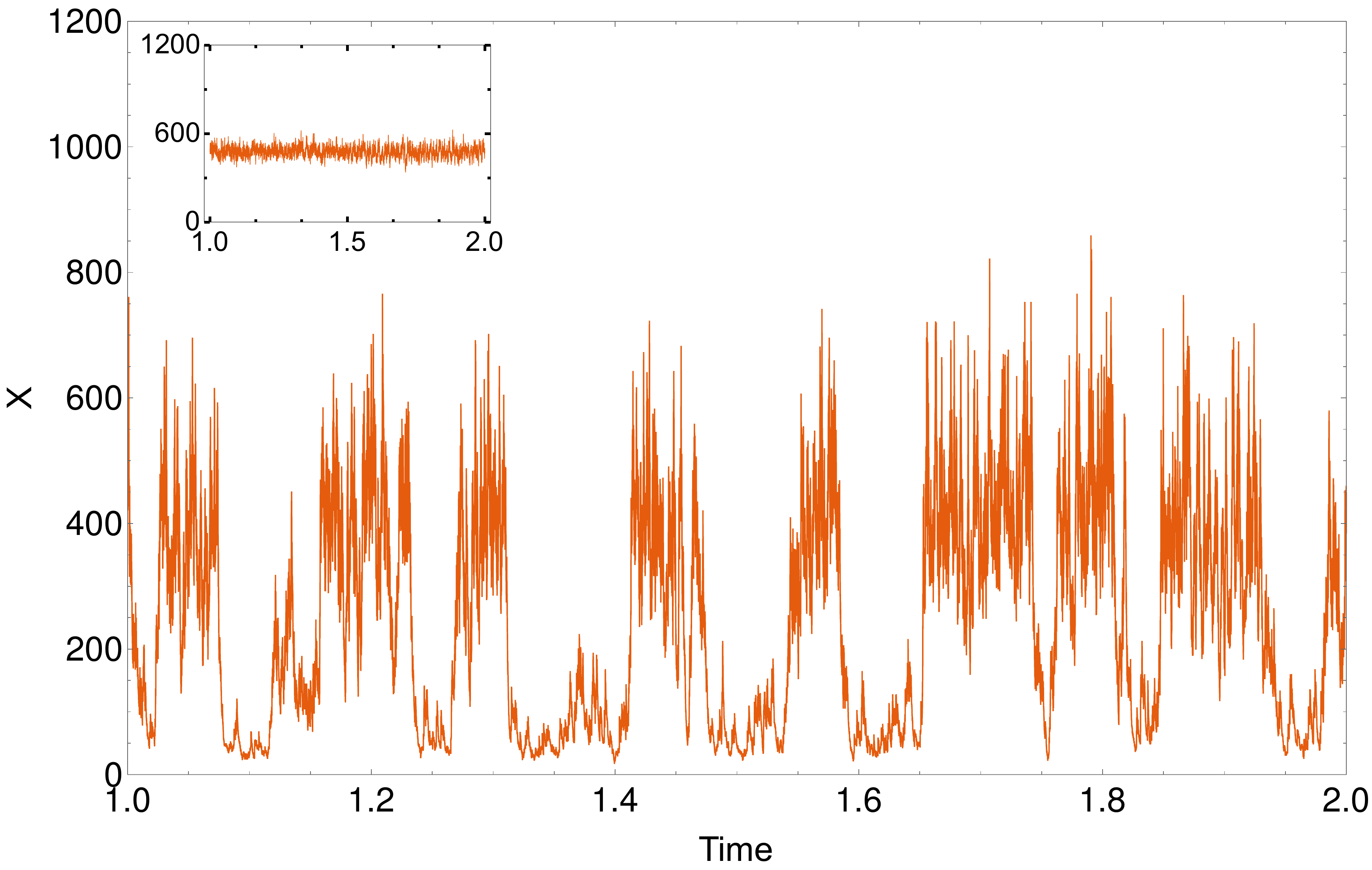}
   \caption{ 
Time evolution of a system with a parameter-set: $k_1 = 3 \cdot 10^6$, $k_2 = 77518$  and $k_3 = 601$. The burst size was set to $r = 25$. The time series shows a bistable characteristic of the system with one stable point at $X=43$ and one at $X =376$. The inset shows the same system without burst ($r = 1$). Here the system has only one stable point at $ X =438$.
\label{fig:series_inlet}}%
\end{figure}

Fig.~\ref{fig:mes_compare} shows the effect of the burst parameter $r$ on the stationary distribution of $x$. All three distributions were generated with the same set of parameters $k_i$, i.e.~with reactions represented by the same deterministic ODE. When the burst size is increased starting from the original model ($r=1$), the distribution changes from one with two maxima to one with one maximum. The reason for this behaviour is that due to the definition of reaction \eqref{burst} an increasing burst size $r$ mainly increases the intrinsic fluctuations for large $X$, i.e., in the region of the right peak. Hence this peak becomes broader and flatter with increasing $r$ much faster than the left one. As in the case $r=1$ shown earlier, there is an excellent agreement between the stationary solution of the CME and the FPE. 
In the following we define the saddle-node bifurcation in the noisy system by the condition that the two extrema merge \cite{bishop_stochastic_2010}. In terms of stochastic bifurcation theory this phenomenon is coined \textit{Phenomenological}(P) Bifurcation, in contrast to \textit{Dynamical}(D) Bifurcations that occur in the fluxes and not the distributions \cite{arnold_random}.
We can investigate the effect of noise strength on the bifurcation. Fig.~\ref{fig:det_compare} shows that noise can destroy the saddle node bifurcation as well as induce it.
  The direction of this shift can be understood from Eq.~\eqref{eq:mes_ext}. Using the implicit function theorem on Eq.~\eqref{eq:mes_ext}, we obtain the following shifts in the location of the maxima and minima on the $k$ axis, 
 \begin{align}
\frac{\partial k}{\partial r} (x, r) &= -\left(\frac{\partial \alpha(x,k,r)}{\partial k}\right)^{-1} \cdot \frac{\partial \alpha(x,k,r)}{\partial r}\nonumber\\
&=\frac{\frac{\partial^2 B(x,k,r)}{\partial x \partial r}}{2\frac{\partial A}{\partial k} - \frac{\partial^2 B}{\partial k \partial x}}\, .\label{eq:bif_shift}
\end{align}
Here, we have used the fact that $A$ does not depend on $r$. 
Inserting for $k$ explicitly the three parameters $\tilde{k}_i$, we obtain
\begin{subequations}
\label{eq:bif_shift_all}
\begin{align}
\frac{\partial \tilde{k}_1}{\partial r}(r, x) &= x \tilde{k}_3\label{eq:bif_shift_k1}\\
\frac{\partial \tilde{k}_2}{\partial r}(r, x) &= -\frac{x}{x-1}\tilde{k}_3\label{eq:bif_shift_k2}\\
\frac{\partial \tilde{k}_3}{\partial r}(r, x) &= \frac{1}{x-r}\tilde{k}_3\, . \label{eq:bif_shift_k3}
\end{align}
\end{subequations}
We note that for $x\gg 1$ Eq.~\eqref{eq:bif_shift_k2} does not depend on $x$, which explains why in Fig.~\ref{fig:det_compare}(b) the distance between two neighboring curves is always identical, and hence why the added noise does not change the qualitative behavior of the system.
Fig.~\ref{fig:crosssect_burst} shows the influence of the parameter $r$ on cross-sections of the  phase diagram Fig.~\ref{fig:number_of_stable}, with features that can be explained from  Eqs.~\eqref{eq:bif_shift_all}. Since $x \gg 0$, the noise induced shifts in $k_2$ direction are according to Eq.~\eqref{eq:bif_shift_k2} equidistant, while the shifts in $k_3$ direction are vanishing according to Eq.~\eqref{eq:bif_shift_k3}. Considering for example the parameter set indicated by the black \textbf{X} in Fig.~\ref{fig:crosssect_burst}, it is easy to understand how a monostable system can become bistable and again monostable with increasing noise.
  In order to check whether the transition from a single to a double peak in the stationary probability distribution (as shown in Fig.~\ref{fig:det_compare} (c) ) is associated with a change from a monostable to a bistable dynamical behavior we generated  stochastic trajectories of the modeled system. We identified parameter values for which a pronounced transition from one to two maxima is observed as the burst size is increased.  As can be seen in the inset in Fig.~\ref{fig:series_inlet},  for the considered parameter values and for $r=1$ the system fluctuates around the single fixed-point at $X=438$. When the burst size is increased to  $r=25$ the system becomes bistable. The time series in Fig.~\ref{fig:series_inlet} shows that indeed the system tends to stay in the vicinity of the maxima at $X=43$ and $X=376$ for some time and switches stochastically between them. This transition to bistability is purely due to increasing burst size (compare Eq.\eqref{burst}), as the change from a mono-stable to a bistable behavior does not occur in the corresponding ODE.

\section{Conclusions}
\label{sec:conclusions}
Our one-dimensional model is the simplest possible system based on mass-action kinetics that can generate bistability due to intrinsic noise. The essential requirement to obtain such a pronounced difference between the deterministic and stochastic version of the dynamics is a bursty production of more than one individual at a time. Since bursty dynamics occurs also in other contexts, our findings are relevant far beyond chemical reaction systems. Bursty noise plays also an important role in  population dynamics. For instance, it was recently shown that the addition of intrinsic noise to the dynamical description can dramatically change the behavior of a species in terms of extinction and survival probability \cite{beer_rare_2016}. Burst noise is also a well known type of noise in semiconductors like e.g. NPN and PNP transistors \cite{knott_characteristics_1978}, where it is know to be responsible for noise spectrum deviations in the low frequency range \cite{zaklikiewicz_influence_1981}.
It can be expected that the effects of intrinsic burst noise become even more important when systems become more complex. However, the study of larger reaction network becomes quickly unfeasible.  While there are several approaches to speed up the simulation of large reaction networks affected by burst noise \cite{bokes_transcriptional_2013,lin_bursting_2016}, our minimal system serves as a good starting point for a theoretical analysis that shapes our basic understanding. 
Since it has been shown that intrinsic noise in 2-dimensional systems can induce oscillations \cite{thomas_signatures_2013,black_stochastic_2012} and create Hopf bifurcations \cite{qian_concentration_2002,ramaswamy_intrinsic_2011} that are not covered by the deterministic description, it can be equally worthwhile to search for  similar minimal models that correctly capture noise-induced bifurcations other than the saddle-node bifurcation.

\section*{Acknowledgments}
JF acknowledges funding by the Hessen State Ministry of Higher Education, Research and the Arts (HMWK) via the ``LOEWE CompuGene'' project.\\
BD was supported in part by Perimeter Institute
of Theoretical Physics. Perimeter Institute is supported
by the Government of Canada through the Department
of Innovation, Science and Economic Development and
by the Province of Ontario through the Ministry of Research, Innovation and Science.


%

\end{document}